# An inferential analysis of the effect of activity based physics instruction on the persistent misconceptions of lecture students

Emily M. Reiser[a)] and Mark E. Markes[b)]

[a)] Formerly, Department of Physics and Physical Science

University of Nebraska-Kearney (deceased)

[b)] Department of Physics and Physical Science

University of Nebraska-Kearney

markesme@unk.edu

PACS:  01.40.Fk, 01.40.Gm

Studies indicate that pre-existing misconceptions negatively impact the effectiveness of traditional physics education. Research has also shown that activity based instruction improves posttest scores on conceptual evaluations. However, the specific effect of activity based instruction on students who have pre-existing misconceptions that lecture instruction fails to remove is an important but unanswered question due to the impossibility of instructing a given student by two different methods. In this paper pre and posttest Force and Motion Conceptual Evaluation results are characterized in terms of wrong to same wrong, wrong to different wrong, wrong to right, etc., and a method of inferring the effect of activity based instruction on a lecture student using the right to wrong responses is presented. Results indicate that a wrong to same wrong response by a lecture student would be more likely to be converted to a wrong to right by activity based instruction than would be a wrong to different wrong response. This result is consistent with cognitive conflict as an aid to learning in activity based classes.

# I. INTRODUCTION

About twenty five years ago researchers began to develop conceptual test instruments to evaluate student understanding of force and motion.[1] As a result of this development several evaluation instruments have emerged.[2-5] Two of the more commonly used instruments are the Force Concept Inventory[6] (FCI) and the Force and Motion Conceptual Evaluation[7] (FMCE). Both the FCI and FMCE underwent similar, lengthy, and rigorous development. In the first stage of this development short answer responses to carefully written conceptual questions, and the results of student interviews, were used to identify many of the student misconceptions associated with a particular question. In the second stage of development these misconceptions were used as the basis for writing “distracters” for multiple choice questions. Thus the presence of an identical wrong answer by a student on a pre and posttest is an indicator of a pre-existing and persistent misconception.[8]

The physics education research community began to systematically study general student use of misconception based distracters about fifteen years ago.[9-14] In 1995, Thornton[15] introduced the term “Conceptual Dynamics” to refer to the phenomenological study of student concepts as a function of time. In Thornton’s method the student responses to multiple choice FMCE questions are mapped into categories called “student views” that are associated with particular beliefs about the relationship between force and motion. Thornton monitored the change in the frequencies describing the use of these views over the course of instruction. More recently, Bao and Redish[16] have introduced a method called “Model Analysis” that is also based on a predetermined set of student models. As in Thornton's analysis the models imply certain beliefs about force and motion on the part of the students. Bao and Redish have applied their method to analyze pre and posttest responses obtain from both the FCI and FMCE.[17]

Thornton's Conceptual Dynamics and Bao and Redish's Model Analysis both involve the application of a predetermined set of models. However, there is one significant aspect of the student's test response that can be addressed without constructing a set of student models: That is, whether or not the student's incorrect posttest response is the same or different from the pretest response. In this paper a study is made of the consistent wrong responses from pre to posttest using the FMCE. For the purpose of this study wrong to same wrong responses will be referred to as "persistent misconceptions". In addition, it has been found possible using this type of characterization to make statistical inferences concerning how activity based instruction might have modified a traditional lecture student's posttest responses. Both the FMCE test results and an analysis of the data will be presented in this paper.

The paper is arranged as follows: In Section II the instruction methods and data parameters are presented. In Section III the results of FMCE testing are presented, and in Section IV the details of the analysis are presented. The paper concludes in Section V with a summary and conclusions.

# II. INSTRUCTION METHODS AND EVALUATION

## A. Instruction Methods

This study involves two instructors and six sections of algebra level introductory physics (one section of each class type each fall semester from 1995 to 1997) at the University of Nebraska-Kearney. Three of these sections employed a Lecture/Demonstration Based (LB) format and were taught by the first instructor. The other three sections employed an Activity Based (AB) format and were taught by the other instructor. Both instructors had over 20 years of teaching experience at the time the data were collected, and both are highly respected teachers who receive excellent student evaluations. (Neither of the authors was directly involved with the instruction.) The LB sections met three 50-minute periods plus one 75-minute period per week with one three-hour laboratory each week. The AB sections met three 110-minute periods plus one 75-minute period per week with an integrated laboratory. Both types of class had a total of 6.75 contact hours per week. There were 87 LB students and 64 AB students. The mean number of LB students per section was 29, and the mean number of AB students per section was 21.

In the LB sections the instruction was via lecture/demonstration, and the LB sections had a required college physics textbook.[18] Problems were assigned from this textbook. The AB sections were based on about 15 to 20 minutes of lecture instruction per 110-minute session with the remainder of the time used by the students, working in groups, to complete sections of a workbook or do experiments. At the time of data collection the AB sections primarily utilized the Calculator Based Laboratory (CBL) technology developed by Texas Instruments. The CBL technology was essentially a TI-85 calculator, various “probes”, and the CBL interface box that connected the calculator to the probes. The activity workbook was written by the instructor who at the time was the principle investigator on a FIPSE grant[19] to integrate activity based methods with the CBL technology. The activity book suggested readings from the textbook. However, it was primarily a stand-alone, guided inquiry workbook organized into activity units focused on particular topics. The workbook was influenced to some extent by early versions of Workshop Physics.[20] However, it was principally an independent work based on the experience of the principle investigator and general findings of the physics education research community. Another primary function of the workbook was to facilitate the integration of the CBL technology into the course. Sample copies of current versions of some activities are available online.[21] The textbook used in the LB sections was also required for the AB sections, but end-of-semester surveys indicated that the students made little use of the text.

**B. Evaluation**

Pre and posttesting was done using an early version of the Force and Motion Conceptual Evaluation[22] (FMCE). The primary difference between the newer version and the version used in this study is that the newer version has more distracters for some of the questions. A total of 87 lecture based students and 64 activity based students were tested. The FMCE has 43 multiple choice questions of which 42 were analyzed.[23] Pretesting was done on both classes sometime during the first two weeks of instruction, and posttesting was done near the end of the semester after all topics on the FMCE had been covered in both classes.

For data recording each student was considered to be a primary sampling unit chosen at random from large hypothetical populations of LB and AB students. Each student $i$ for class type $k$ ($k = 0$ for LB instruction, $k = 1$ for AB instruction) was associated with the 42 questions of the FMCE, and the number of pretest wrong $N_w^{(k)}(i)$ and pretest right $N_r^{(k)}(i)$ responses were related as

$$N_r^{(k)}(i) + N_w^{(k)}(i) = 42 . \qquad (1)$$

When both the pretest and the posttest are considered, there are two possible outcomes for pretest right questions, right to right ($rr$) and right to wrong ($rw$). For pretest wrong questions there are three possible outcomes: wrong to same wrong ($ww$), wrong to different wrong ($ww'$), and wrong to right ($wr$). Each student $i$ is associated with five random variables $n_{rr}^{(k)}(i)$, $n_{rw}^{(k)}(i)$, $n_{ww}^{(k)}(i)$, $n_{ww'}^{(k)}(i)$, and $n_{wr}^{(k)}(i)$, which are the total numbers of outcomes (frequencies) in the five response categories for student $i$.

For the purpose of comparing the numbers of persistent misconceptions (the $ww$ responses) only the pretest wrong questions are of interest. However, different students have different numbers of pretest wrong responses, and this must be considered in the analysis. In order to take this variation into account the three possible pretest wrong outcomes ($ww$, $ww'$, $wr$) are used to calculate three random variables per student $x_{ww}^{(k)}(i)$, $x_{ww'}^{(k)}(i)$, and $x_{wr}^{(k)}(i)$, as

$$x_{ww}^{(k)}(i) = n_{ww}^{(k)}(i) / N_w^{(k)}(i) , \qquad (2a)$$

$$x_{ww'}^{(k)}(i) = n_{ww'}^{(k)}(i) / N_w^{(k)}(i) , \qquad (2b)$$

and

$$x_{wr}^{(k)}(i) = n_{wr}^{(k)}(i) / N_w^{(k)}(i) . \qquad (2c)$$

For simplicity, Eqs. (2) can be written

$$x_\rho^{(k)}(i) = n_\rho^{(k)}(i) / N_w^{(k)}(i), \tag{3}$$

where $\rho =$ $ww$, $ww'$, or $wr$. For the pretest wrong questions the LB and AB students are characterized by the means of these three variables summed over the students of the two class types

$$\overline{x}_\rho^{(k)} = \frac{1}{N_s^{(k)}} \sum_{i=1}^{N_s^{(k)}} x_\rho^{(k)}(i). \tag{4}$$

Here $N_s^{(k)}$ is the total number of students of class type $k$. The quantities $\overline{x}_\rho^{(k)}$, are the estimated mean probabilities for a student who answered a question wrong on the pretest to answer that same question wrong ($\rho = ww$), different wrong ($\rho = ww'$), or right ($\rho = wr$) on the posttest.

Similarly, the pretest right responses of student $i$ from class type $k$ are characterized by

$$x_\rho^{(k)}(i) = n_\rho^{(k)}(i) / N_r^{(k)}(i), \tag{5}$$

where $\rho =$ $rw$ or $rr$, and the mean probabilities per student are again given by Eq. (4) with $\rho = rw$ or $\rho = rr$.

One complication that must be considered in the analysis is that the quantities $x$ are ratios of random variables instead of simple random variables. However, Hansen et. al.[24] have shown that ratios of random variables can be consider to be simple random variables under three conditions: (1) the sample is a simple random sample of elementary units; (2) the denominator is the number of units in a specific class of the sample; and (3) the numerator is an aggregate value for some characteristic of the units enumerated in the denominator. In this analysis the quantities $x$ satisfy these requirements and will be considered to be simple random variables.

## III. FMCE RESULTS

Figure 1 shows a histogram of the number of pretest wrong responses per student grouped in four bins: 0-24, 25-29, 30-34, and 35-39. The picket heights were determined by first counting the total number of pretest wrongs for each student and then placing the student into the appropriate bin. The bin values were then normalized by dividing by the total number of students in each class type (LB or AB). Thus, to determine the total number of LB students who had between 0 and 24 wrong answers on the pretest one would multiply the bin's value (0.08 for this bin) by the number of LB students (87), or $0.08 \times 87 = 7$. The numbers of students in the other bins are calculated in an analogous way.

From Fig. 1 it can be seen that the bin with the largest number of students for both types of class is the 30-34 bin. The number of LB students in this bin is $0.60 \times 87 = 52$ and the number of AB students in this bin is $0.66 \times 64 = 42$. It can be noted that there is on average a fairly good match between the fractions of students in each bin. A chi-square test for independence yielded $p = 0.20$ which weakly suggests there may be a minor dependence on class type. This could be due to either the different class types attracting slightly different groups of students or to the time at which the pretest was administered. The LB sections typically had the pretest given the second week of instruction, and the AB sections had it typically given the first week of instruction. It should be noted, however, that a slight dissimilarity between the two student groups should not significantly influence the analysis results because all reported $x$ values are a result of normalization against either the total number of pretest right or pretest wrong questions for each student. In general, the overall pretest results are very consistent with the findings of other researchers who have found that very predictably an average student will miss about 75 percent of the questions on the pretest.[25]

Figure 2 shows the mean probabilities $\overline{x}_{ww}$, $\overline{x}_{ww'}$, $\overline{x}_{wr}$, $\overline{x}_{rr}$, and $\overline{x}_{rw}$ for the LB and AB sections. For $ww$, $ww'$, and $wr$ these are the sample based mean probabilities that a pretest wrong question will be answered same wrong, different wrong, or right respectively on the posttest. For $rw$ and $rr$ these are the sample based mean probabilities that a pretest right question will be answered wrong or right respectively on the posttest. The error bars are set at plus or minus one standard deviation. It can be seen that the most significant differences are found in the wrong to same wrong ($ww$) and the wrong to right ($wr$) categories with AB instruction significantly reducing the probability of a $ww$ response and significantly increasing the probability of $wr$ response. The wrong to different wrong response ($ww'$) and the pretest right responses ($rw, rr$) are affected less significantly.

# IV. ANALYSIS

The analysis is based on the concept of taking a large hypothetical population of $N$ students through the operations shown schematically in Fig. 3. The actual students participating in the study are assumed to be random samples taken from this population after posttesting. These samples are indicated by small boxes within the Population 0 and Population 1 boxes.

As shown in Fig. 3 the students are first pretested and then given LB instruction followed by posttesting. The students after LB instruction and posttesting are labeled Population 0. Mean probabilities $\bar{X}_{\rho}^{(k)}$ for the populations can be defined in analogy with the mean sample probabilities defined in Section II

$$\bar{X}_{\rho}^{(k)} = \frac{1}{N}\sum_{i=1}^{N} X_{\rho}^{(k)}(i) \tag{6}$$

where $\rho = ww$, $ww'$, $wr$, $rw$, or $rr$. In this and the following section sample based quantities will be indicated using lower case Roman symbols and population based quantities will be indicated by upper case Roman symbols. After the posttest the entire population of $N$ students is hypothetically assumed to be returned to their preinstruction state, reinstructed using AB methods, and posttested a second time and relabeled Population 1. Since the reinitialization is assumed to remove all memory of the LB instruction, the AB students tested in practice can be considered as a random sample taken from Population 1. Further, if $N$ is much larger than the sample sizes, sampling with replacement can be assumed even though, in practice, LB students cannot appear in the AB sample.

The changes due to reinitialization and reinstruction using activity based methods can be written

$$\Delta\bar{X}_{\rho} = \bar{X}_{\rho}^{(1)} - \bar{X}_{\rho}^{(0)}, \tag{7}$$

where the $\bar{X}_{\rho}^{(k)}$ are mean probabilities per student in Population $k$. Thus they must satisfy

$$\Delta\bar{X}_{rw} + \Delta\bar{X}_{rr} = 0 \tag{8}$$

and

$$\Delta\bar{X}_{ww} + \Delta\bar{X}_{ww'} + \Delta\bar{X}_{wr} = 0. \tag{9}$$

Equations (8) and (9) imply that the effect of reinitialization and reinstruction of the lecture students by activity based methods can be represented by transfers of probability $A$, $B$, $C$, and $D$ as shown in Fig. 4. The transfers resulting from a hypothetical reinstruction using activity based methods can be expressed as

$$D = -\Delta\bar{X}_{rw}, \tag{10}$$

$$A + C = \Delta\bar{X}_{wr}, \tag{11}$$

and

$$A + B = -\Delta\bar{X}_{ww}. \tag{12}$$

Given $A$, $B$, $C$, and $D$ the effect of a hypothetical reinstruction of the population of $N$ lecture based students is determined. However, without additional information, Eqs. (11) and (12) cannot be solved for $A$, $B$, and $C$ because there are more unknowns than independent equations.

A solution is made possible by an additional assumption that relates the values $C$ and $D$. Recent interviews[26] of students who answered right to wrong on certain FMCE questions have indicated that they generally could not state specifically why they were able to answer a question correctly on the pretest and yet failed to answer it correctly on the posttest. A typical response was that the pretest answer must have been a "lucky guess". Another possibility, although none of the interviewed students mentioned it, is that the pretest right response might have been obtained in spite of faulty physical understanding. An examination of the FMCE reveals that this is possible for at least some of the questions.[27] Given that the pretest right response was fortuitous, or made in spite of a faulty physical model, right to wrong responses are essentially equivalent to wrong to different wrong responses, and there must be some relationship between $C$ and $D$.

This relationship can be obtained as follows. First express $C$ and $D$ as

$$C = \frac{1}{N}\sum_i \frac{n^{(0)}_{ww'}(i)}{N^{(0)}_w(i)} \frac{n^{(0)}_{ww',wr}(i)}{n^{(0)}_{ww'}(i)} = \frac{1}{N}\sum_i X^{(0)}_{ww'}(i)\, f_{ww',wr}(i) \qquad (13a)$$

and

$$D = \frac{1}{N}\sum_i \frac{n^{(0)}_{rw}(i)}{N^{(0)}_r(i)} \frac{n^{(0)}_{rw,rr}(i)}{n^{(0)}_{rw}(i)} = \frac{1}{N}\sum_i X^{(0)}_{rw}(i)\, f_{rw,rr}(i)\,, \qquad (14a)$$

where $n^{(0)}_{ww',wr}(i)$ and $n^{(0)}_{rw,rr}(i)$ are the numbers of questions transferred from $ww'$ to $wr$ and from $rw$ to $rr$ by hypothetical reinstruction of lecture student $i$. The factors $f_{ww',wr}(i) = n^{(0)}_{ww',wr}(i)/n^{(0)}_{ww'}(i)$ and $f_{rw,rr}(i) = n^{(0)}_{rw,rr}(i)/n^{(0)}_{rw}(i)$ are the "transition probabilities" for a question to be transferred from $ww'$ to $wr$ and from $rw$ to $rr$ respectively for student $i$. Analogous expressions can be written for $A$ and $B$. Assuming the transition probabilities $f_{ww',wr}(i)$ and $f_{rw,rr}(i)$ are uncorrelated with $X^{(0)}_{ww'}(i)$ and $X^{(0)}_{rw}(i)$, Eqs. (13a) and (14a) can be written

$$C = \left( \frac{1}{N} \sum_i X_{ww'}^{(0)}(i) \right) \left( \frac{1}{N} \sum_i f_{ww',wr}(i) \right) = \overline{X}_{ww'}^{(0)} \, \overline{f}_{ww',wr} \,, \qquad (13b)$$

and

$$D = \left( \frac{1}{N} \sum_i X_{rw}^{(0)}(i) \right) \left( \frac{1}{N} \sum_i f_{rw,rr}(i) \right) = \overline{X}_{rw}^{(0)} \, \overline{f}_{rw,rr} \,. \qquad (14b)$$

Based on the assumed equivalence of the $ww'$ and $rw$ responses as discussed above, the mean transition probabilities $\overline{f}_{ww',wr}$ and $\overline{f}_{rw,rr}$ can be set equal, and Eqs. (13b), (14b), and (10) yield

$$C = -\frac{\overline{X}_{ww'}^{(0)}}{\overline{X}_{rw}^{(0)}} \Delta \overline{X}_{rw} \,. \qquad (15)$$

In practice only sample based quantities are available while Eqs. (10) – (12) and (15) refer to populations. However, sample based values can be used to estimate population based values to within a certain estimated confidence interval determined from the standard deviation of the sample, and standard deviations can be calculated for products, quotients, and sums given the standard deviations of the factors or terms. The expressions for the standard deviation of products, quotients, and sums of sample mean values are derived in the Appendix. The probability transfer $C$ is estimated as

$$C \approx -\left( \frac{\overline{x}_{ww'}^{(0)}}{\overline{x}_{rw}^{(0)}} \right) \Delta \overline{x}_{rw} \,, \qquad (16)$$

where $\Delta \overline{x}_{rw} = \overline{x}_{rw}^{(1)} - \overline{x}_{rw}^{(0)}$, and the standard deviation of $C$ is calculated in two steps. First estimates of the standard deviations are calculated for the factors appearing in Eq. (16) using the sample standard deviations given in Tables I and II and Eqs. (A10) and (A13) from the Appendix

$$\sigma_{\Delta\bar{x}_{rw}} \approx \sqrt{s^2_{\bar{x}^{(1)}_{rw}} + s^2_{\bar{x}^{(0)}_{rw}}} \tag{17}$$

and

$$\sigma_{\bar{x}^{(0)}_{ww'}/\bar{x}^{(0)}_{rw}} \approx \sqrt{\left(\frac{\bar{x}^{(0)}_{ww'}}{\bar{x}^{(0)}_{rw}}\right)^2 \left[\frac{s^2_{\bar{x}^{(0)}_{ww'}}}{\left(\bar{x}^{(0)}_{ww'}\right)^2} + \frac{s^2_{\bar{x}^{(0)}_{rw}}}{\left(\bar{x}^{(0)}_{rw}\right)^2}\right]} . \tag{18}$$

Here, and in the remainder of this section, the population based standard deviations $\sigma$ are estimated using sample based standard deviations as $s^2_{\bar{x}^{(k)}_p} = s^2_{x^{(k)}_p} / N^{(k)}_s$. The standard deviation of $C$ can now be estimated using Eq. (A6)

$$\sigma_C \approx \sqrt{\left(\bar{x}^{(0)}_{ww'} / \bar{x}^{(0)}_{rw}\right)^2 \sigma^2_{\Delta\bar{x}_{rw}} + \left(\Delta\bar{x}_{rw}\right)^2 \sigma^2_{\bar{x}^{(0)}_{ww'}/\bar{x}^{(0)}_{rw}}} . \tag{19}$$

The mean transition probability $\bar{f}_{ww',wr}$ is estimated from Eq. (13b)

$$\bar{f}_{ww',wr} \approx C / \bar{x}^{(0)}_{ww'} , \tag{20}$$

with the standard deviation estimated using Eq. (A10)

$$\sigma_{\bar{f}_{ww',wr}} \approx \sqrt{\left(\frac{C}{\bar{x}^{(0)}_{ww'}}\right)^2 \left[\frac{\sigma^2_C}{C^2} + \frac{s^2_{\bar{x}^{(0)}_{ww'}}}{\left(\bar{x}^{(0)}_{ww'}\right)^2}\right]} . \tag{21}$$

Once $C$ is determined $A$ can be estimated using Eq. (11)

$$A \approx \left(\bar{x}^{(1)}_{wr} - \bar{x}^{(0)}_{wr}\right) - C , \tag{22}$$

with the standard deviation obtained by applying Eq. (A13) twice

$$\sigma_A \approx \sqrt{s^2_{\bar{x}^{(1)}_{wr}} + s^2_{\bar{x}^{(0)}_{wr}} + \sigma^2_C} . \tag{23}$$

Analogous to Eqs. (13) and (14) the transition probability associated with $A$ is

$$\bar{f}_{ww,wr} \approx A / \bar{x}^{(0)}_{ww} , \tag{24}$$

and the standard deviation of $\overline{f}_{ww,wr}$ is

$$\sigma_{\overline{f}_{ww,wr}} \approx \sqrt{\left(\frac{A}{\overline{x}_{ww}^{(0)}}\right)^2 \left[\frac{\sigma_A^2}{A^2} + \frac{s_{\overline{x}_{ww}^{(0)}}^2}{\left(\overline{x}_{ww}^{(0)}\right)^2}\right]}. \tag{25}$$

The probability transfer $B$ is estimated using Eq. (12)

$$B \approx -\left(\overline{x}_{ww}^{(1)} - \overline{x}_{ww}^{(0)}\right) - A, \tag{26}$$

with the standard deviation obtained by applying Eq. (A13) twice

$$\sigma_B \approx \sqrt{s_{\overline{x}_{ww}^{(1)}}^2 + s_{\overline{x}_{ww}^{(0)}}^2 + \sigma_A^2} . \tag{27}$$

The transition probability associated with $B$ is

$$\overline{f}_{ww,ww'} \approx B / \overline{x}_{ww}^{(0)} , \tag{28}$$

and the standard deviation of this quantity is

$$\sigma_{\overline{f}_{ww,ww'}} \approx \sqrt{\left(\frac{B}{\overline{x}_{ww}^{(0)}}\right)^2 \left[\frac{\sigma_B^2}{B^2} + \frac{s_{\overline{x}_{ww}^{(0)}}^2}{\left(\overline{x}_{ww}^{(0)}\right)^2}\right]}. \tag{29}$$

Numerical values obtained using Tables I and II and Eqs. (16) – (29) are given in Table III.

## V. SUMMARY AND CONCULSIONS

This paper has presented a method, based on sampling theory, that infers the effect that activity based instruction would have on traditional lecture students were it possible to return them to their initial knowledge state and reinstruct them using activity based methods. This is done by categorizing responses on FMCE pre and posttests as wrong to same wrong ( $ww$ ), wrong to different wrong ( $ww'$ ), wrong to right ( $wr$ ), right to wrong ( $rw$ ), or right to right ( $rr$ ). In addition, two assumptions are made: the first assumption (based on student interviews) is that the $ww'$ response is equivalent to the $rw$ response, and the second assumption is that the transition probabilities $f_{\rho,\rho'}$ (the probability that an LB question in the $\rho$ response category will be transformed to the $\rho'$ category by hypothetical AB reinstruction) vanish for directions opposite the arrows in Fig. 4. The central theoretical object of the analysis is a single large hypothetical population of $N$ students that is initially instructed using traditional lecture methods. The lecture students actually participating in the study are assumed to be a random sample taken from this population which is labeled Population 0. After posttesting Population 0 is assumed to be returned to its preinstruction state and reinstructed using activity based methods. After reinstruction and posttesting a second time, the population is labeled Population 1. The activity based students actually participating in the study are assumed to be a random sample taken from Population 1.

The primary quantities of interest are the mean transition probabilities $f_{\rho,\rho'}$ with $\rho$ and $\rho'$ = $ww$, $ww'$, or $wr$. The assumed equivalence of the $ww'$ and $wr$ responses is used to justify the assumption $\overline{f}_{ww',wr} = \overline{f}_{rw,rr}$ which allows all mean transition probabilities to be calculated. It is found that $\overline{f}_{ww,wr}$ and $\overline{f}_{ww',wr}$ are resolved at one standard deviation with the estimated value of $\overline{f}_{ww,wr}$ about twice the estimated value of $\overline{f}_{ww',wr}$. The transition probability $\overline{f}_{ww,ww'}$ is not completely resolved from $\overline{f}_{ww',wr}$. However, $\overline{f}_{ww,ww'}$ is very small, and the estimated value of $\overline{f}_{ww',wr}$ is about five times as large as the estimated value of $\overline{f}_{ww,ww'}$. Generally, the results suggest that $\overline{f}_{ww,wr} > \overline{f}_{ww',wr} > \overline{f}_{ww,ww'}$ with $\overline{f}_{ww,ww'}$ significantly smaller than either $\overline{f}_{ww,wr}$ or $\overline{f}_{ww',wr}$.

The practical impossibility of returning lecture students to their initial condition demands a discussion of some underlying cognitive theory in which this kind of analysis would be plausible. For an example of a cognitive theory under which this analysis would NOT be plausible, consider the point of view that students are essentially “blank slates” where learning is a “recording” process acting on the slate. In this case the analysis presented in this paper is not plausible because reinitialization erases all the slates and all questions become equivalent. However, it is likely that learning is at least partly a constructive process that makes use of a student’s pre-existing knowledge. From the constructivist[28,29] point of view the results presented in this paper suggest that a lecture student who answers wrong to same wrong on a particular question has a working model that can be applied to that question, and hence also has many of the primitive cognitive building blocks[30] (diSessa’s “p-prims”) from which an expert physical model can be constructed. It then follows that activity based instruction is effective at stimulating those students to use these mental elements to construct expert knowledge, and it supports the idea that cognitive conflict is a viable and beneficial aspect of an activity based classroom.

## ACKNOWLEDGMENTS

The authors wish to acknowledge useful suggestions and comments from Dr. Robert J. Whitaker, Dept. of Physics, Astronomy and Material Science, Southwest Missouri State University, Dr. Teara Archwamety, Dept. of Counseling and School Psychology, University of Nebraska-Kearney, and Dr. C. Trecia Markes, Dept. of Physics and Physical Science, University of Nebraska-Kearney. They also express their gratitude to Ms. Shea Holman for spreadsheet development and data entry, to Mrs. Karen Malmkar for conducting the student interviews, to The Fund for the Improvement of Post Secondary Education grant P116B51449 under which the data was collected, and the University of Nebraska-Kearney, University Research and Creative Activity grant which funded this work.

## APPENDIX

In this appendix approximate expressions for $\sigma^2_{\bar{x}\,\bar{y}}$ and $\sigma^2_{\bar{x}/\bar{y}}$ will be derived as well as an exact expression for $\sigma^2_{\bar{x}\pm\bar{y}}$. Let samples of size $n$ with means $\bar{x}_\alpha$ and $\bar{y}_\beta$ be taken from populations of size $N$ with means $\bar{X}$ and $\bar{Y}$. If there are $M$ ways of taking samples from each of the populations then the variance of $\bar{x}_\alpha \bar{y}_\beta$ is[31]

$$\sigma^2_{\bar{x}\,\bar{y}} = \frac{1}{M^2}\sum_{\alpha=1}^{M}\sum_{\beta=1}^{M}\left(\bar{x}_\alpha \bar{y}_\beta - \bar{X}\,\bar{Y}\right)^2 . \qquad \text{(A1)}$$

Now define $\Delta\bar{x}_\alpha$ and $\Delta\bar{y}_\beta$ as

$$\bar{x}_\alpha = \bar{X} + \Delta\bar{x}_\alpha \qquad \text{(A2)}$$

and

$$\bar{y}_\beta = \bar{Y} + \Delta\bar{y}_\beta . \qquad \text{(A3)}$$

To first order in $\Delta\bar{x}_\alpha / \bar{X}$ and $\Delta y_\beta / \bar{Y}$

$$\bar{x}_\alpha \bar{y}_\beta - \bar{X}\,\bar{Y} = \bar{Y}\Delta\bar{x}_\alpha + \bar{X}\Delta y_\beta \qquad \text{(A4)}$$

and

$$\sigma^2_{\bar{x}\,\bar{y}} = \frac{\bar{Y}^2}{M}\sum_{\alpha=1}^{M}\Delta\bar{x}^2_\alpha + \frac{\bar{X}^2}{M}\sum_{\beta=1}^{M}\Delta y^2_\beta + 2\frac{\bar{X}\,\bar{Y}}{M^2}\sum_{\alpha=1}^{M}\sum_{\beta=1}^{M}\Delta\bar{x}_\alpha \Delta y_\beta . \qquad \text{(A5)}$$

For uncorrelated samples the last term vanishes, and the variance is

$$\sigma^2_{\bar{x}\,\bar{y}} = \bar{Y}^2\sigma^2_{\bar{x}} + \bar{X}^2\sigma^2_{\bar{y}} . \qquad \text{(A6)}$$

For $\bar{x}/\bar{y}$ the variance is

$$\sigma^2_{\bar{x}/\bar{y}} = \frac{1}{M^2}\sum_{\alpha=1}^{M}\sum_{\beta=1}^{M}\left(\frac{\bar{x}_\alpha}{\bar{y}_\beta} - \frac{\bar{X}}{\bar{Y}}\right)^2 , \qquad \text{(A7)}$$

with $\Delta\bar{x}_\alpha$ and $\Delta\bar{y}_\beta$ defined as in Eqs. (A2) and (A3). For $\Delta\bar{x}_\alpha \ll \bar{X}$ and $\Delta\bar{y}_\beta \ll \bar{Y}$

$$\frac{\bar{x}_\alpha}{\bar{y}_\beta} = \frac{\bar{X}}{\bar{Y}}\left[1+\frac{\Delta\bar{x}_\alpha}{\bar{X}}\right]\left[1-\frac{\Delta\bar{y}_\beta}{\bar{Y}}+\left(\frac{\Delta\bar{y}_\beta}{\bar{Y}}\right)^2-\cdots\right]. \qquad \text{(A8)}$$

To second order in $\Delta\bar{x}_\alpha/\bar{X}$ and $\Delta\bar{y}_\beta/\bar{Y}$

$$\frac{1}{M^2}\sum_{\alpha=1}^{M}\sum_{\beta=1}^{M}\left(\frac{\bar{x}_\alpha}{\bar{y}_\beta}-\frac{\bar{X}}{\bar{Y}}\right)^2 = \left(\frac{\bar{X}}{\bar{Y}}\right)^2\left\{\frac{1}{M}\sum_{\alpha=1}^{M}\left(\frac{\Delta\bar{x}_\alpha}{\bar{X}}\right)^2+\frac{1}{M}\sum_{\beta=1}^{M}\left(\frac{\Delta\bar{y}_\beta}{\bar{Y}}\right)^2-2\sum_{\alpha=1}^{M}\sum_{\beta=1}^{M}\frac{\Delta\bar{x}_\alpha}{\bar{X}}\frac{\Delta\bar{y}_\beta}{\bar{Y}}\right\}. \qquad \text{(A9)}$$

For uncorrelated samples the last term vanishes, and the variance is

$$\sigma^2_{\bar{x}/\bar{y}} = \left(\frac{\bar{X}}{\bar{Y}}\right)^2\left\{\frac{\sigma^2_{\bar{x}}}{\bar{X}^2}+\frac{\sigma^2_{\bar{y}}}{\bar{Y}^2}\right\}. \qquad \text{(A10)}$$

For $\bar{x} \pm \bar{y}$ the variance is

$$\sigma^2_{\bar{x}\pm\bar{y}} = \frac{1}{M^2}\sum_{\alpha=1}^{M}\sum_{\beta=1}^{M}\left[\left(\bar{x}_\alpha \pm \bar{y}_\beta\right)-\left(\bar{X}\pm\bar{Y}\right)\right]^2. \qquad \text{(A11)}$$

Equations (A2) and (A3) yield

$$\sigma^2_{\bar{x}\pm\bar{y}} = \frac{1}{M}\sum_{\alpha=1}^{M}\Delta\bar{x}_\alpha^2+\frac{1}{M}\sum_{\beta=1}^{M}\Delta\bar{y}_\beta^2 \pm \frac{2}{M^2}\sum_{\alpha=1}^{M}\sum_{\beta=1}^{M}\Delta\bar{x}_\alpha\Delta y_\beta . \qquad \text{(A12)}$$

For uncorrelated samples the last term vanishes, and the variance is

$$\sigma^2_{\bar{x}\pm\bar{y}} = \sigma^2_{\bar{x}}+\sigma^2_{\bar{y}}. \qquad \text{(A13)}$$

[a] Formerly, Department of Physics and Physical Science, University of Nebraska-Kearney, (deceased).

[b] Electronic mail: markesme@unk.edu

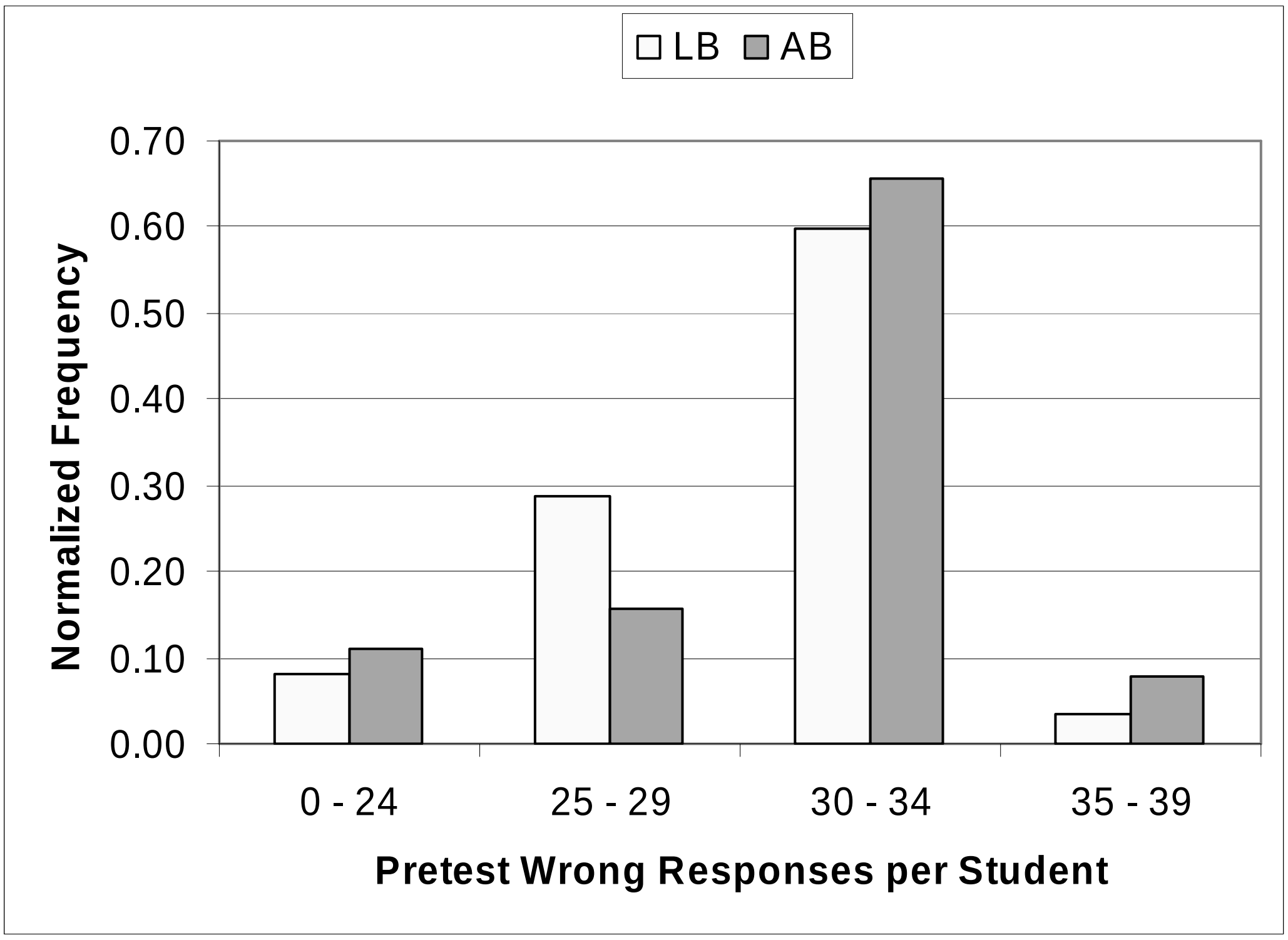

**Fig. 1. Comparison of pretest wrong responses for LB and AB students. Chi square $p = 0.20$.**

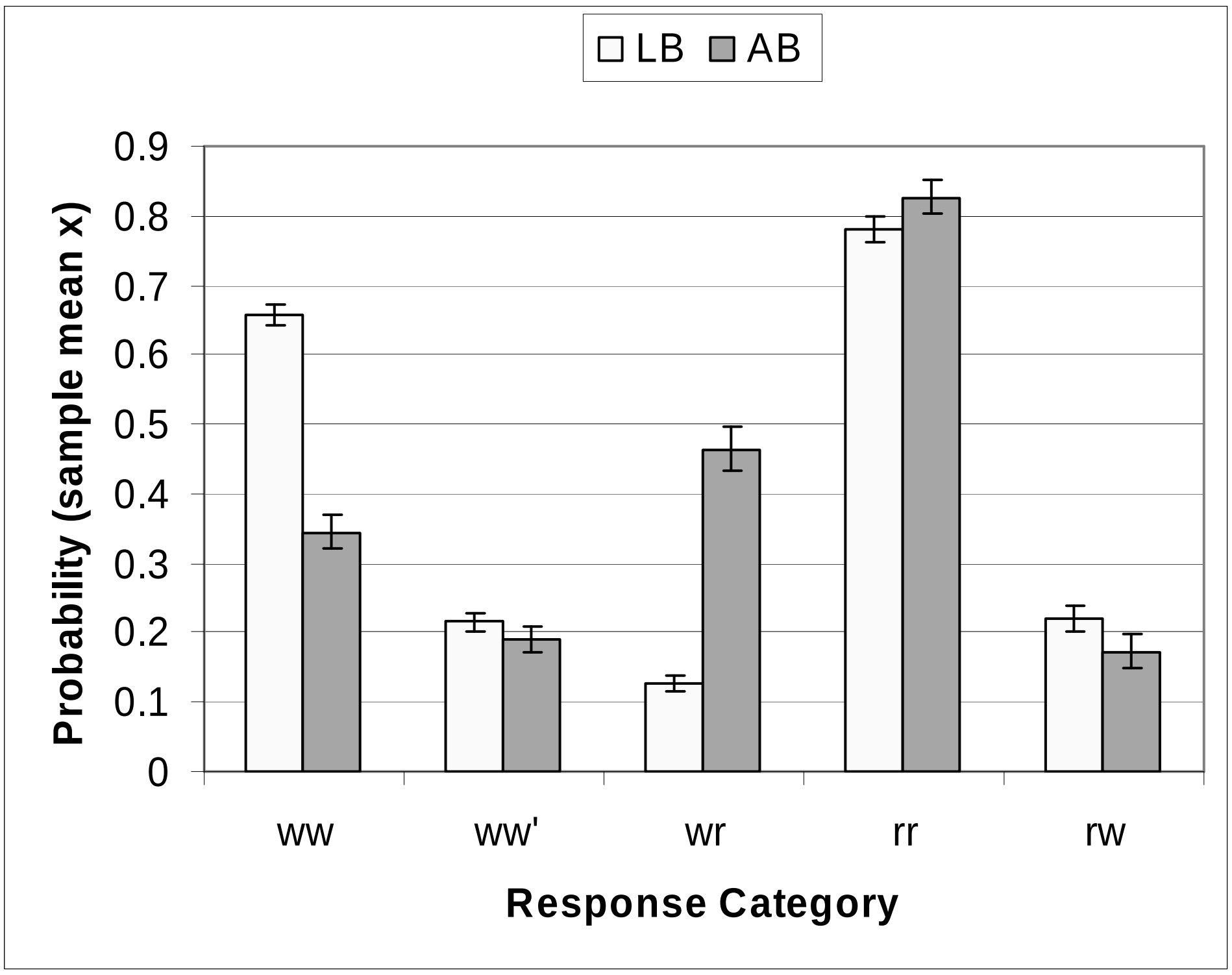

**Fig. 2. Mean probabilities x for the LB and AB students. Error bars are at plus and minus one standard deviation.**

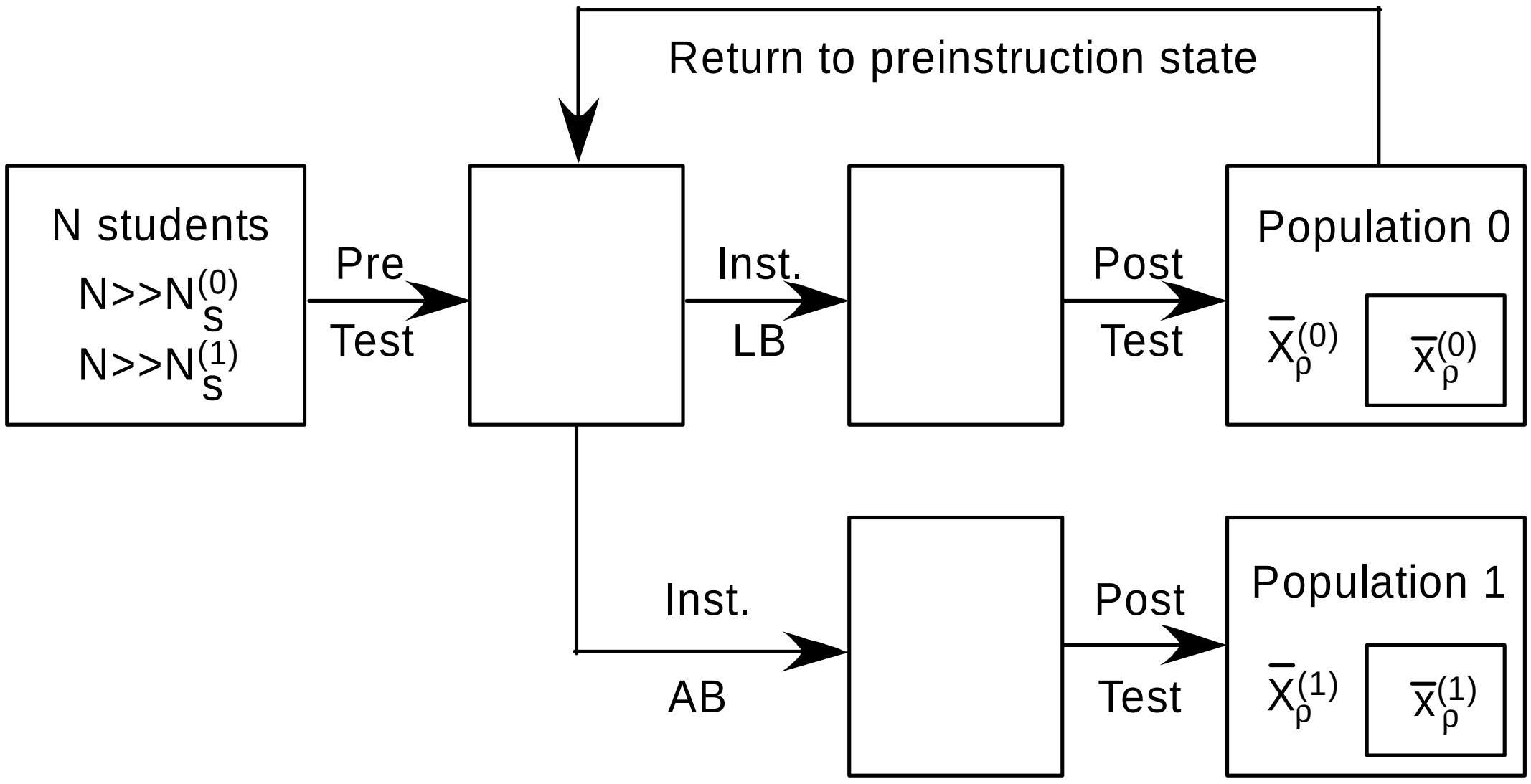

**Fig. 3. A schematic representation of the hypothetical population *N* used in the analysis, the actual student samples, and the hypothetical reinstruction.**

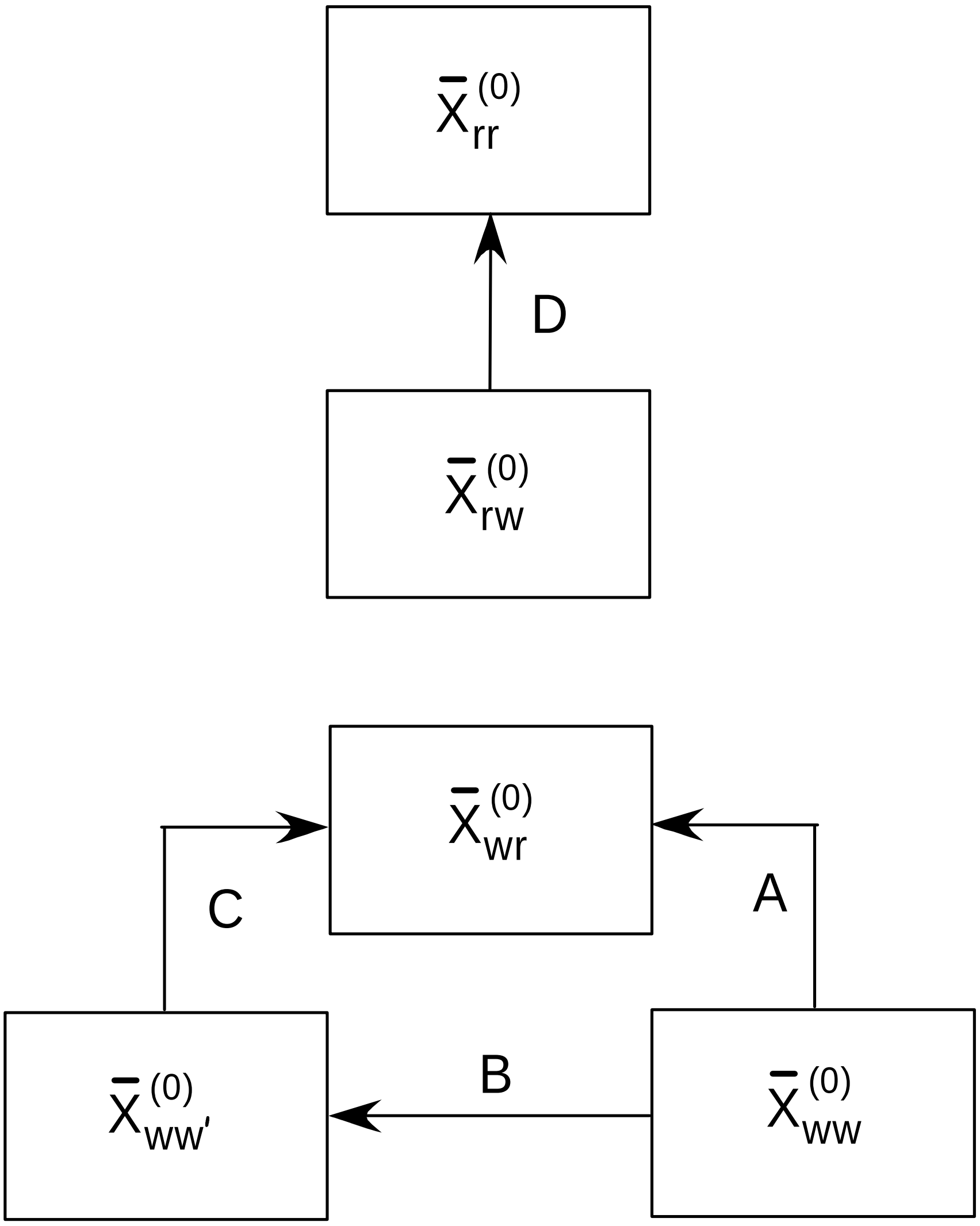

**Fig. 4. The effect of activity based instruction on the population of lecture based students.**

**Table I. The $x$ parameters for the pretest right responses.**

| Method | No. Pretest Right $N_r^{(k)}$ | $\overline{x}_{rr}$ | (s.d. of $\overline{x}_{rr}$) | $\overline{x}_{rw}$ | (s.d. of $\overline{x}_{rw}$) |
|---|---|---|---|---|---|
| Lecture Based $N_s = 87$, $k = 0$ | 1090 | 0.779 | (0.019) | 0.221 | (0.019) |
| Activity Based $N_s = 64$, $k = 1$ | 737 | 0.827 | (0.025) | 0.173 | (0.025) |

**Table II. The $x$ parameters for the pretest wrong responses.**

| Method | No. Pretest Wrong $N_w^{(k)}$ | $\overline{x}_{ww}$ (s.d. of $\overline{x}_{ww}$) | $\overline{x}_{ww'}$ (s.d. of $\overline{x}_{ww'}$) | $\overline{x}_{wr}$ (s.d. of $\overline{x}_{wr}$) |
|---|---|---|---|---|
| Lecture Based $N_s = 87$, $k = 0$ | 2564 | 0.658 (0.015) | 0.215 (0.012) | 0.127 (0.011) |
| Activity Based $N_s = 64$, $k = 1$ | 1951 | 0.345 (0.023) | 0.191 (0.018) | 0.464 (0.031) |

**Table III. The probability transfers ( $A$ , $B$ , $C$ ) and transition probabilities $A / \bar{x}_{ww}^{(0)}$ $B / \bar{x}_{ww}^{(0)}$ , and $C / \bar{x}_{ww'}^{(0)}$ . Standard deviations are given in parenthesis.**

| | $A$ (s.d.)<br>$A/\bar{x}_{ww}^{(0)}$ (s.d.) | | $B$ (s.d.)<br>$B/\bar{x}_{ww}^{(0)}$ (s.d.) | | $C$ (s.d.)<br>$C/\bar{x}_{ww'}^{(0)}$ (s.d.) | |
|---|---|---|---|---|---|---|
| Probability Tranfers | 0.290 | (0.045) | 0.023 | (0.053) | 0.047 | (0.031) |
| Transition Probabilities | 0.441 | (0.069) | 0.035 | (0.080) | 0.219 | (0.145) |